# HARMONIC AND TIMBRE ANALYSIS OF TABLA STROKES

Anirban Patranabis[1], Kaushik Banerjee[1], Vishal Midya[2], Sneha Chakraborty[2], Shankha Sanyal[1], Archi Banerjee[1], Ranjan Sengupta[1] and Dipak Ghosh[1]

[1]Sir C V Raman Centre for Physics and Music, Jadavpur University, Kolkata 700032, India
[2]Indian Statistical Institute, Kolkata, India

**Abstract**

Indian twin drums mainly bayan and dayan (tabla) are the most important percussion instruments in India popularly used for keeping rhythm. It is a twin percussion/drum instrument of which the right hand drum is called dayan and the left hand drum is called bayan. Tabla strokes are commonly called as `bol', constitutes a series of syllables. In this study we have studied the timbre characteristics of nine strokes from each of five different tablas. Timbre parameters were calculated from LTAS of each stroke signals. Study of timbre characteristics is one of the most important deterministic approach for analyzing tabla and its stroke characteristics. Statistical correlations among timbre parameters were measured and also through factor analysis we get to know about the parameters of timbre analysis which are closely related. Tabla strokes have unique harmonic and timbral characteristics at mid frequency range and have no uniqueness at low frequency ranges.

**Introduction**

Among the percussion instruments, 'tabla' is one of the most important musical instruments in India. Tabla plays an important role in accompanying vocalists, instrumentalists and dancers in every style of music from classical to light in India, mainly used for keeping rhythm. The 'right hand' drum, called the dayan (also called the dahina, or the tabla) is a conical (almost cylindrical) drum shell carved out of a solid piece of hard wood. The 'open' end is covered by a composite membrane. The 'left hand' drum, called the bayan (also called the duggi) is a hemispherical bowl shaped drum made of polished copper, brass, bronze, or clay. Both of them have an 'open' end, covered by a composite membrane. The drum head (puri), is unique, and is made of goatskin, the lao. There is a weight in the middle, the syahi or gub. The syahi is perfect circle, in the middle of the puri, it is a semi-permanent paste made of coal dust, iron fillings, and rice paste. Around the outside of the puri, is a ring of thicker skin, this is called the chanti, this is not attached to the lao. The puri is laced by buffalo skin straps, baddhi, and tensioned by round wooden 'chocks', called gittak.

As an accompanying instrument, Tabla serves the purpose of keeping rhythm by repeating a theka (beat-pattern) and adorns the vocal/instrumental music that it is accompanying. In this music, the choice of strokes is precise, each one functioning like a note in a melody; the timbral and rhythmic structures are equally important and carefully integrated into a singing line.

Tabla strokes are typically inharmonic in nature but strongly pitched resonant strokes (Raman 1934). The sounds of most drums are characterized by strongly inharmonic spectra; however tablas, especially the dayan are an exception. This was pointed out as early as 1920 by C. V. Raman and S. Kumar. Raman further refined

the study in a later paper (Ghosh R N, 1922, Raman C V, 1934, Rao K N, 1938). Thereafter several theoretical and experimental studies were held on the dynamics of the instrument (Ramakrishna B S, 1957, Sarojini T et. al, 1958, Banerjee B M et. al, 1991, Courtney D, 1999). The classical model put forth by Raman represents the sound of tabla-dayan, as having a spectrum consisting of five harmonics; these are the fundamental with its four overtones (Courtney D, 1999). Here we studied the timbre characteristics of tabla strokes.

**Strokes chosen for analysis**

Tabla playing has a very well developed formal structure and an underlying "language" for representing its sounds. A tabla `bol's constitutes a series of syllables which correlate to the various strokes of the tabla. Here we have considered nine tabla strokes. Stroke ‘Ta/Na’ executes by lightly pressing the ring finger down in order to mute the sound while index finger strikes the edge. Stroke ‘Ti’ executes by striking the dayan on the 2$^{nd}$ circle with the index finger and by keeping the finger on that position causes more damping but after striking if the index finger release quickly to give an open tone it produces ‘Teen’. Stroke ‘Ghe’ executes by striking the bayan with middle and index finger keeping the wrist on the membrane but after striking if released quickly it produces ‘Ge’. Stroke ‘Thun’ executes by striking on the centre circle of dayan with index, middle, ring and little fingers together and by quickly releasing. Stroke ‘Tu’ executes by striking at the corner of centre circle of dayan with index finger only and immediately after striking finger will lift. Stroke ‘Te’ executes by striking the dayan with middle and ring finger at the centre of the circle. Stroke ‘Re’ executes by striking the dayan with index finger at the centre of the circle and by keeping the finger on that position causes more damping.

**Experimental procedure**

All the strokes were played by eminent tabla players and the sound was recorded in a noise free acoustic room. Membrane of tabla 1, 2 and 3 have diameter 5”, tabla 4 has a diameter 5.5” and tabla 5 has a diameter 6”. We have 5x9 = 45 stroke signals. Each of these sound signals was digitized with sample rate of 44.1 kHz, 16 bit resolution and in a mono channel. All the sound samples are of same length. We used Long Term Average Spectrum (LTAS) for timbre analysis. A rigorous statistical analysis was done based on Principal Component Analysis and Varimax with Kaiser Normalization.

**Timbre analysis**

Timbre is defined in ASA (1960) as that quality which distinguishes two sounds with the same pitch, loudness and duration. This definition defines what timbre is not, not what timbre is. Timbre is generally assumed to be multidimensional, where some of the dimensions have to do with the spectral envelope, the time envelope, etc. Many timbre parameters have been proposed to encompass the timbre dimensions. Among all timbre parameters important parameters are irregularity, tristimulus1 (T1), tristimulus 2 (T2), tristimulus 3 (T3), odd and even parameters, spectral centroid and brightness (Park T H 2004, Grey J M 1977, Patranabis A 2011). Beside these we also measured pitch, attack time, difference in frequency and amplitude between two highest peaks of LTAS and average RMS power of each signals (Sengupta R. et. a., 2004). From the LTAS of each signal above mentioned timbre features were measured of which some are harmonic and some perceptual features.

From figure 1 to 7 it is observed that stroke 'ta' for all tablas have low brightness hence this stroke possess lower energy for all tablas. Since this stroke executes by striking index finger at the edge and such process of stroke cause weak resonance in the cavity of tabla. Brightness of all other strokes is different for five tablas. So timbre variations are confirmed in five tablas. Brightness and hence energy of all the nine strokes are high for the 3rd tabla. So it may be assumed that the resonance takes place in 3rd tabla is the highest among other tablas. But brightness and hence energy of all the nine strokes are low for the 4th and 5th tablas. So it may be assumed that the resonance takes place in 4th and 5th tablas are the lowest among other tablas. Strokes 'thun', ‘ti’ and ‘re’ for all tablas have high centroid hence this stroke is of high pitched for all the tablas. Since these strokes executes by striking at the vicinity of the circle and such process of stroke cause strong resonance in the cavity of tabla. Centroid of all other strokes is different for five tablas. Tristimulus 1 for stroke ‘ta’ is high for 4th and 5th tablas, while tristimulus 2 for ‘tu’, ‘teen’ and ‘ghe’ are high for 4th and 5th tablas. Both the tablas show lower tristimulus 3. Besides these strokes all other strokes have lower fundamental and also less energized lower partials, energy pumps up at higher partials. Comparing all strokes it is observed that irregularity among partials are higher for tabla 3 and 4 for the strokes ‘ghe’, ‘tu’ and ‘teen’. No significant difference is observed in odd and even parameters. Stroke 'Ge' of tabla 1 is different from others viz. brightness and centroid both are low and stronger 2nd, 3rd and 4th harmonics and low irregularity. Other tablas show uniformity in timbre for the stroke 'Ge'. Strokes 'Ghe', 'tu','teen' and 'ta' of tabla 3 and 4 are different from others viz. brightness and centroid are too low and stronger 2nd, 3rd and 4th harmonics and higher irregularity, while other three tablas show uniformity and all these four strokes are free stroke. For stroke 'thun', tabla 3 and 4 differ from other tablas only in brightness i.e. its CG of amplitude. Stroke 'ti' of tabla 3 is different from others viz. brightness and centroid both are low, stronger 2nd, 3rd and 4th harmonics and higher irregularity. For strokes 're' and 'te', tabla 3 and 4 differs from other tablas only in irregularity, in which both the strokes are made at the centre circle and both are damped strokes. Damped strokes have higher brightness and spectral centroid than the free strokes. So this concludes the fact that style (nature and intensity) of strokes of player of 1st tabla is different than others. Also style of strokes is similar for the players of tabla 3 and 4.

From table 2, it is observed that tristimulus 2 (T2) and tristimulus 3 (T3) are highly correlated while tristimulus 1 (T1) is weakly correlated with T2 and T3. This concludes the fact that fundamental (corresponds to T1) of different tabla strokes are different. Also fundamental is weak compared to its harmonics, while mid and higher frequency partials behave similarly. Odd and even harmonics are equally proportionate and are highly correlated to each other and so tabla strokes are harmonically good to hear. T2 and T3 both have high correlation with irregularity and spectral centroid. This concludes the fact that high frequency partials have higher order of irregularity among partials. Also brightness (i.e. centre of gravity of amplitude) and spectral centroid (i.e. centre of gravity of frequency) are highly correlated.

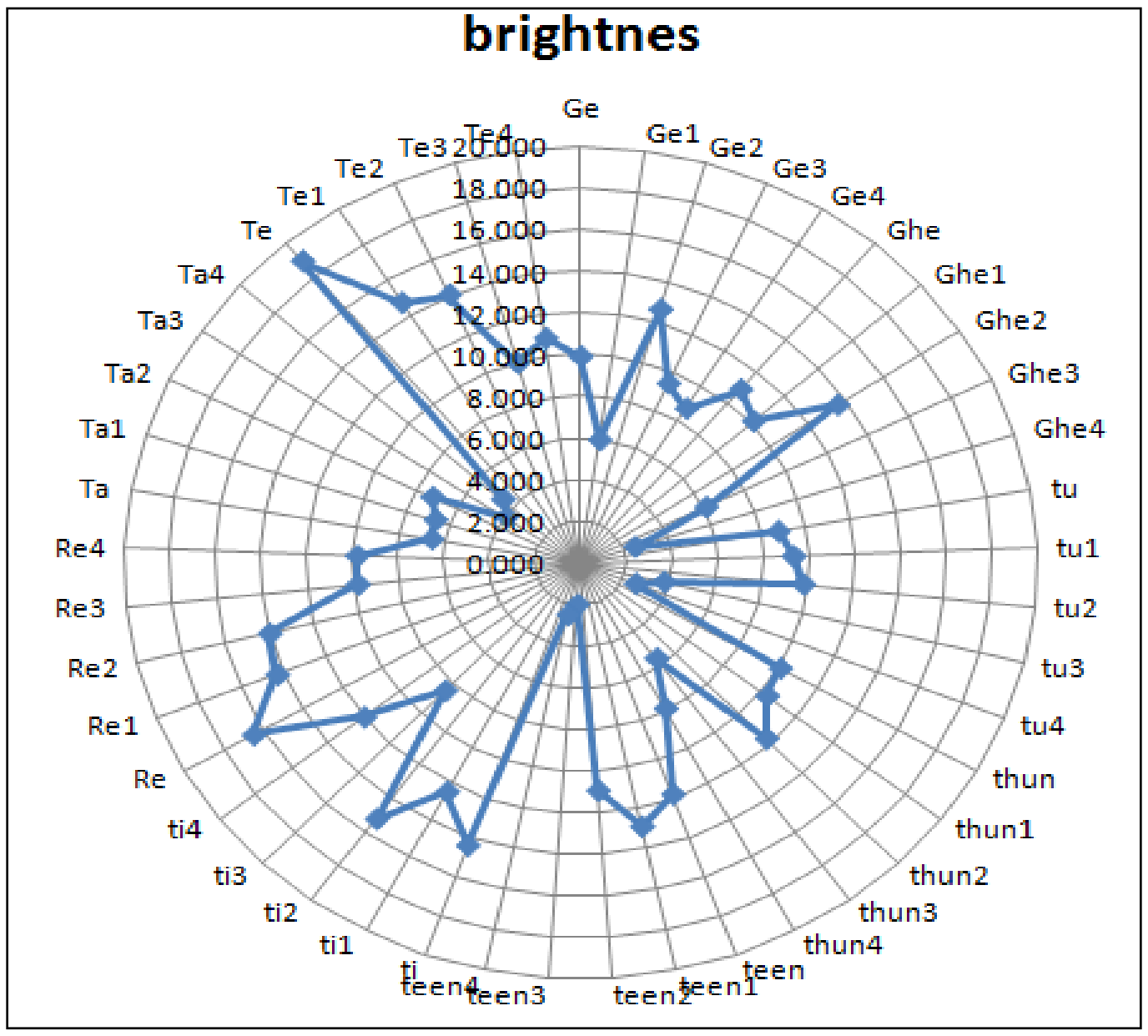

Fig. 1: Variation of brightness

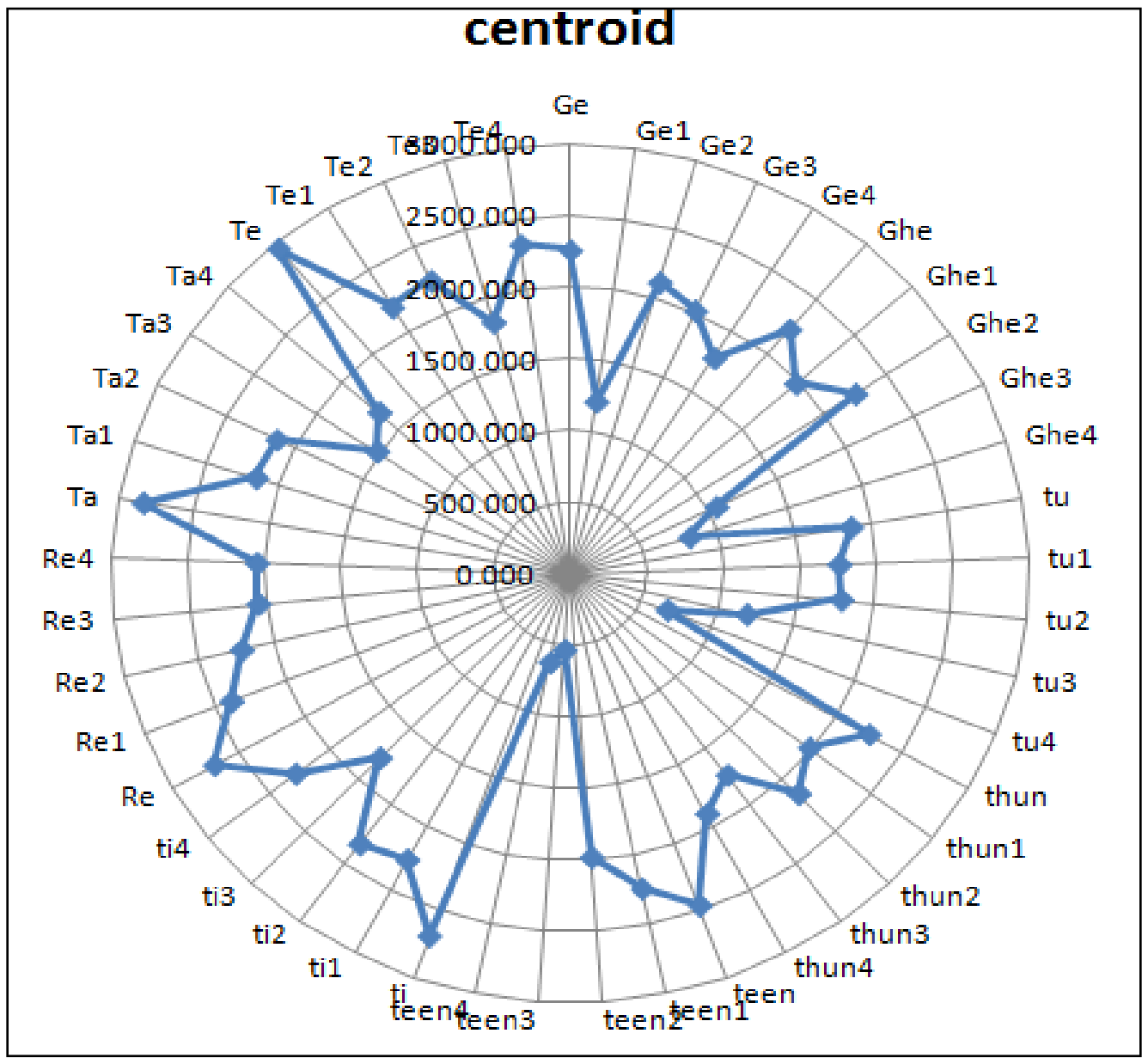

Fig. 2: Variation of centroid

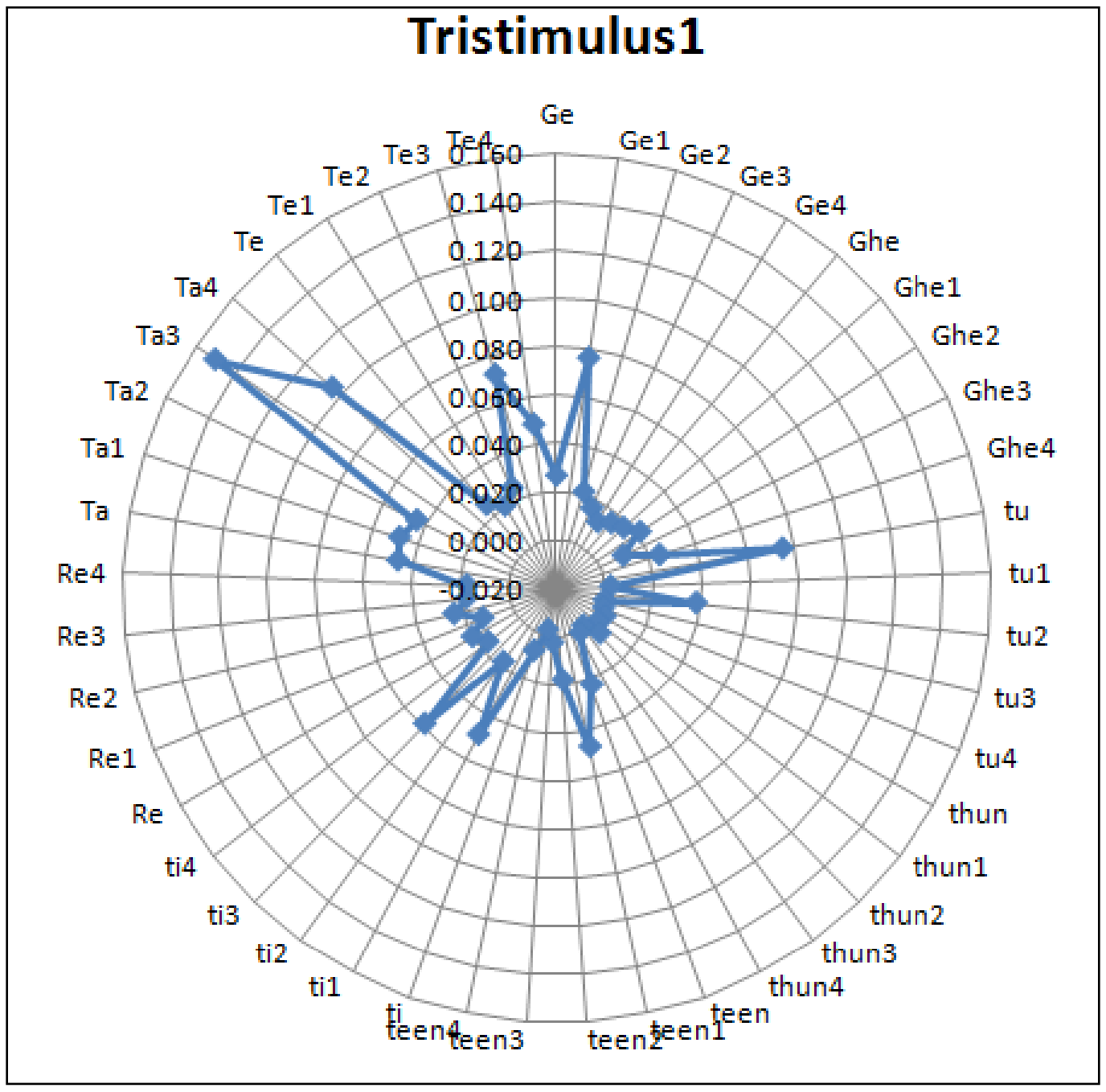

Fig. 3: Variation of tristimulus 1

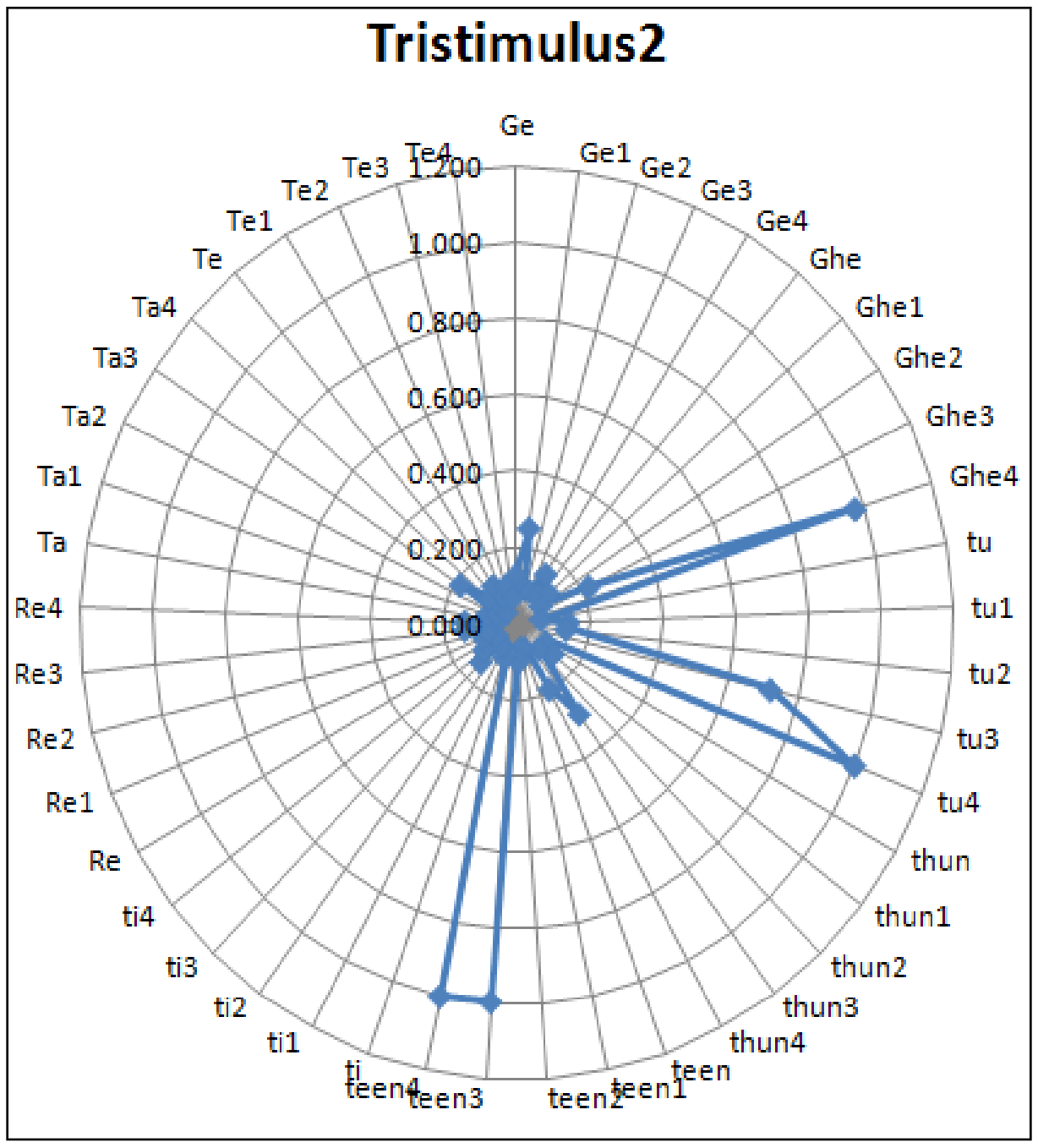

Fig. 4: Variation of tristimulus 2

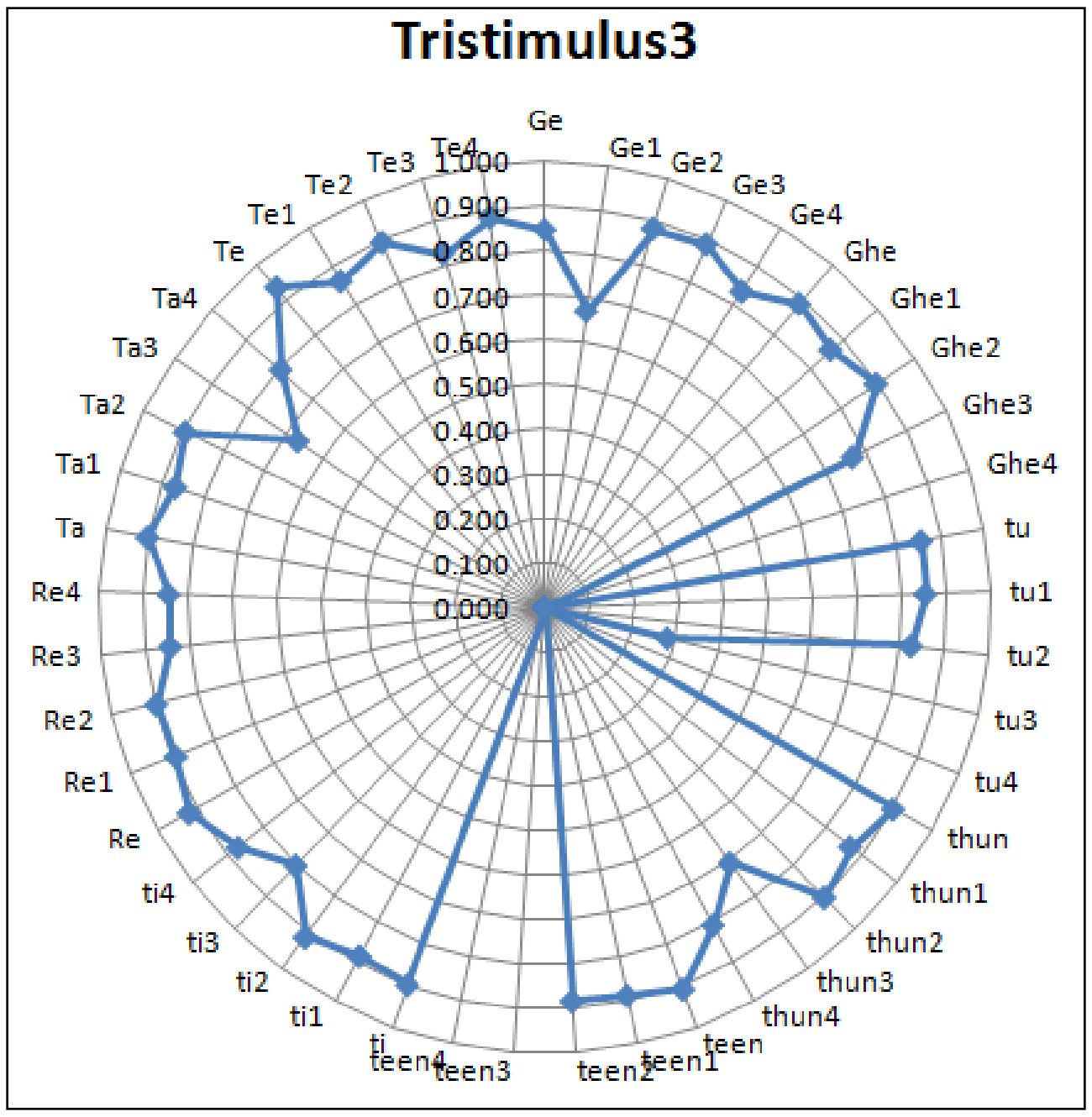

**Fig. 5: Variation of tristimulus 3**

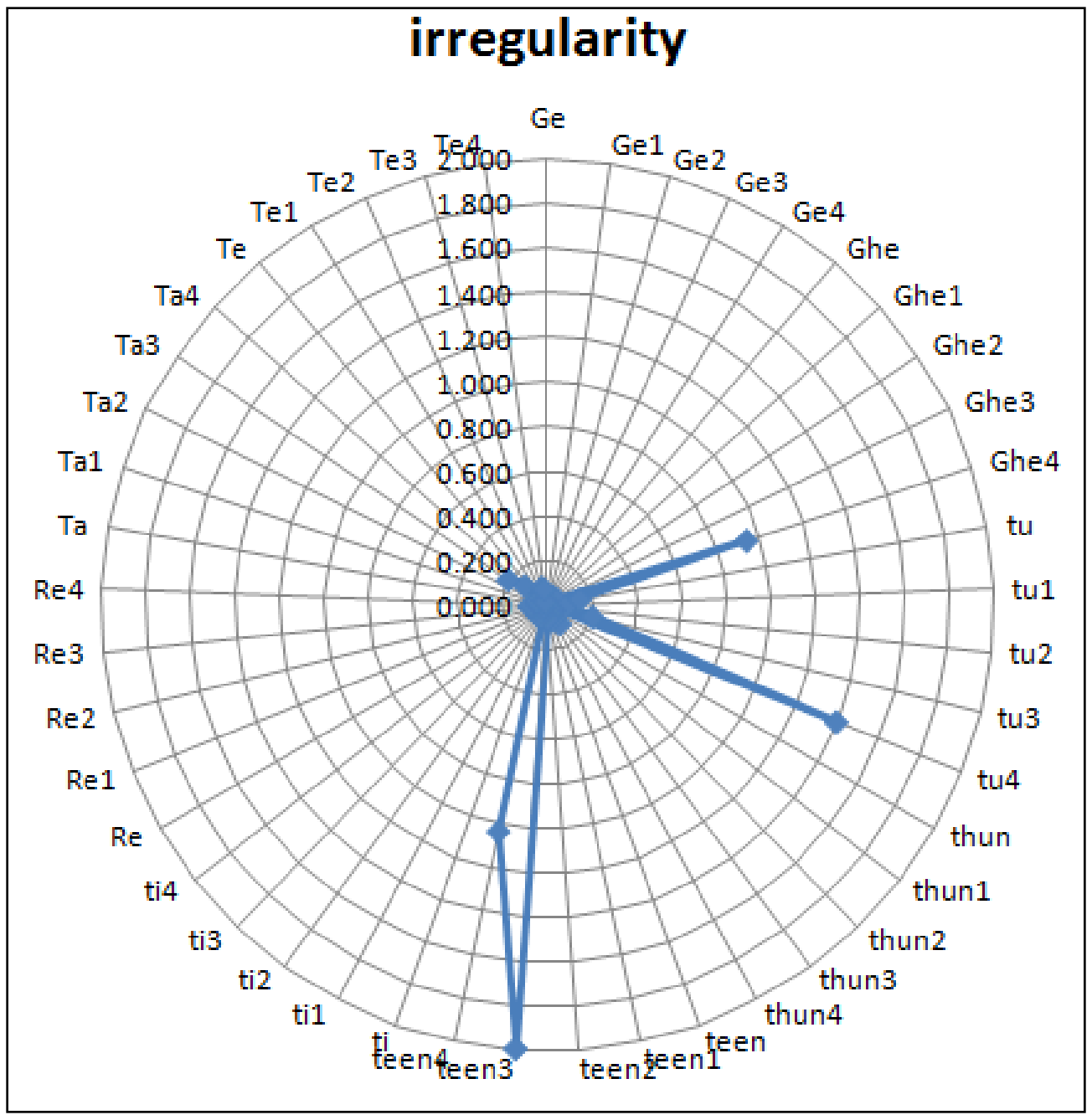

**Fig. 6: Variation of irregularity**

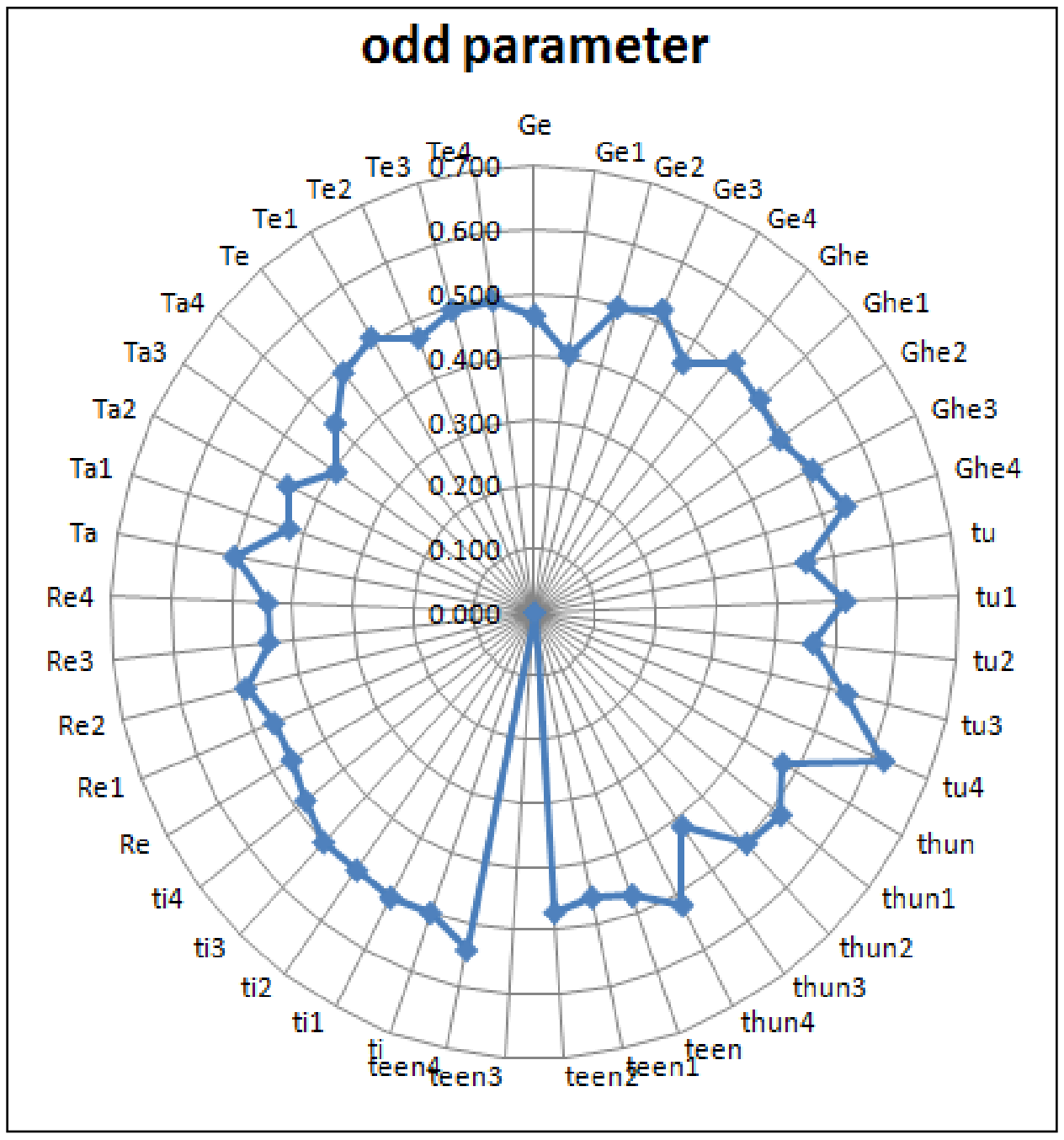

**Fig. 7: Variation of odd parameter**

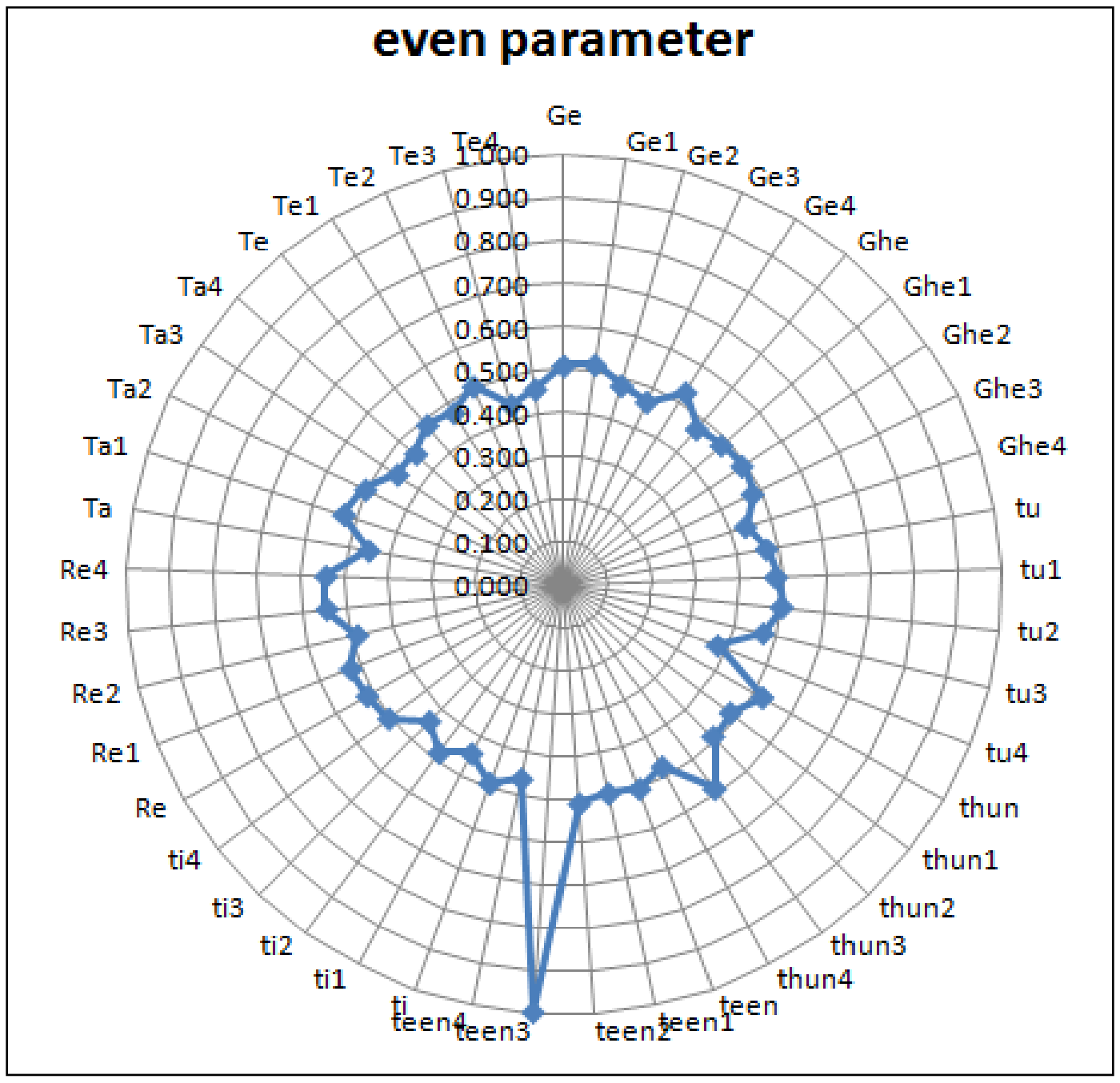

**Fig. 8: Variation of even parameter**

## Table 2. Correlation coefficients of various timbre parameters

| Correlations Between different Timbre parameters | | | | | | | | | | | | | | | |
|---|---|---|---|---|---|---|---|---|---|---|---|---|---|---|---|
| | | Brightness | Tristimulus1 | Tristimulus2 | Tristimulus 3 | Odd parameter | Even Parameter | Spectral Irregularity | Spectral inharmonicity | Spectral Centroid | pitch | Attack time | Average RMS power | Diff in freq of 2 peaks | Diff in amp of 2 peaks |
| Brightness | Pearson Correlation | **1** | **.472** | **-.572** | **.482** | **-.228** | **-.318** | **-.911**** | **-.269** | **.851**** | **-.169** | **.439** | **-.628** | **.111** | **.138** |
| | Sig. (2-tailed) | | **.238** | **.139** | **.226** | **.587** | **.442** | **.002** | **.519** | **.007** | **.689** | **.277** | **.095** | **.794** | **.744** |
| | N | **8** | **8** | **8** | **8** | **8** | **8** | **8** | **8** | **8** | **8** | **8** | **8** | **8** | **8** |
| Tristimulus1 | Pearson Correlation | **.472** | **1** | **-.793*** | **.386** | **-.817*** | **-.073** | **-.285** | **.034** | **.527** | **-.181** | **.177** | **-.046** | **.044** | **.167** |
| | Sig. (2-tailed) | **.238** | | **.019** | **.346** | **.013** | **.864** | **.495** | **.937** | **.179** | **.667** | **.674** | **.913** | **.918** | **.692** |
| | N | **8** | **8** | **8** | **8** | **8** | **8** | **8** | **8** | **8** | **8** | **8** | **8** | **8** | **8** |
| Tristimulus2 | Pearson Correlation | **-.572** | **-.793*** | **1** | **-.868**** | **.492** | **.328** | **.229** | **.144** | **-.687** | **-.055** | **-.430** | **-.185** | **.140** | **.044** |
| | Sig. (2-tailed) | **.139** | **.019** | | **.005** | **.215** | **.427** | **.586** | **.734** | **.060** | **.897** | **.288** | **.661** | **.741** | **.917** |
| | N | **8** | **8** | **8** | **8** | **8** | **8** | **8** | **8** | **8** | **8** | **8** | **8** | **8** | **8** |
| Tristimulus3 | Pearson Correlation | **.482** | **.386** | **-.868**** | **1** | **-.079** | **-.439** | **-.115** | **-.247** | **.611** | **.231** | **.507** | **.317** | **-.247** | **-.203** |
| | Sig. (2-tailed) | **.226** | **.346** | **.005** | | **.853** | **.276** | **.785** | **.556** | **.108** | **.582** | **.199** | **.444** | **.555** | **.629** |
| | N | **8** | **8** | **8** | **8** | **8** | **8** | **8** | **8** | **8** | **8** | **8** | **8** | **8** | **8** |
| Oddparameter | Pearson Correlation | **-.228** | **-.817*** | **.492** | **-.079** | **1** | **-.516** | **.165** | **-.460** | **-.210** | **.498** | **.293** | **.145** | **.039** | **.057** |
| | Sig. (2-tailed) | **.587** | **.013** | **.215** | **.853** | | **.191** | **.696** | **.251** | **.617** | **.209** | **.481** | **.732** | **.927** | **.894** |
| | N | **8** | **8** | **8** | **8** | **8** | **8** | **8** | **8** | **8** | **8** | **8** | **8** | **8** | **8** |
| EvenParameter | Pearson Correlation | **-.318** | **-.073** | **.328** | **-.439** | **-.516** | **1** | **.147** | **.733*** | **-.435** | **-.581** | **-.786*** | **-.168** | **-.129** | **-.344** |
| | Sig. (2-tailed) | **.442** | **.864** | **.427** | **.276** | **.191** | | **.728** | **.039** | **.281** | **.131** | **.021** | **.691** | **.760** | **.403** |
| | N | **8** | **8** | **8** | **8** | **8** | **8** | **8** | **8** | **8** | **8** | **8** | **8** | **8** | **8** |
| SpectralIrregularity | Pearson Correlation | -.911$^{**}$ | -.285 | .229 | -.115 | .165 | .147 | 1 | .257 | -.677 | .260 | -.177 | .829$^{*}$ | -.225 | -.286 |
| | Sig. (2-tailed) | .002 | .495 | .586 | .785 | .696 | .728 | | .538 | .065 | .534 | .675 | .011 | .591 | .493 |
| | N | 8 | 8 | 8 | 8 | 8 | 8 | 8 | 8 | 8 | 8 | 8 | 8 | 8 | 8 |
| SpectralInharmonicity | Pearson Correlation | -.269 | .034 | .144 | -.247 | -.460 | .733$^{*}$ | .257 | 1 | -.234 | -.794$^{*}$ | -.227 | -.013 | -.553 | -.353 |
| | Sig. (2-tailed) | .519 | .937 | .734 | .556 | .251 | .039 | .538 | | .577 | .019 | .588 | .975 | .155 | .390 |
| | N | 8 | 8 | 8 | 8 | 8 | 8 | 8 | 8 | 8 | 8 | 8 | 8 | 8 | 8 |
| SpectralCentroid | Pearson Correlation | .851$^{**}$ | .527 | -.687 | .611 | -.210 | -.435 | -.677 | -.234 | 1 | .003 | .561 | -.317 | .093 | .367 |
| | Sig. (2-tailed) | .007 | .179 | .060 | .108 | .617 | .281 | .065 | .577 | | .994 | .148 | .444 | .827 | .371 |
| | N | 8 | 8 | 8 | 8 | 8 | 8 | 8 | 8 | 8 | 8 | 8 | 8 | 8 | 8 |
| pitch | Pearson Correlation | -.169 | -.181 | -.055 | .231 | .498 | -.581 | .260 | -.794$^{*}$ | .003 | 1 | .170 | .368 | .604 | .135 |
| | Sig. (2-tailed) | .689 | .667 | .897 | .582 | .209 | .131 | .534 | .019 | .994 | | .687 | .369 | .113 | .749 |
| | N | 8 | 8 | 8 | 8 | 8 | 8 | 8 | 8 | 8 | 8 | 8 | 8 | 8 | 8 |
| attacktime | Pearson Correlation | .439 | .177 | -.430 | .507 | .293 | -.786$^{*}$ | -.177 | -.227 | .561 | .170 | 1 | .001 | -.127 | .028 |
| | Sig. (2-tailed) | .277 | .674 | .288 | .199 | .481 | .021 | .675 | .588 | .148 | .687 | | .998 | .764 | .947 |
| | N | 8 | 8 | 8 | 8 | 8 | 8 | 8 | 8 | 8 | 8 | 8 | 8 | 8 | 8 |
| averageRMSpower | Pearson Correlation | -.628 | -.046 | -.185 | .317 | .145 | -.168 | .829$^{*}$ | -.013 | -.317 | .368 | .001 | 1 | -.372 | -.053 |
| | Sig. (2-tailed) | .095 | .913 | .661 | .444 | .732 | .691 | .011 | .975 | .444 | .369 | .998 | | .365 | .900 |
| | N | 8 | 8 | 8 | 8 | 8 | 8 | 8 | 8 | 8 | 8 | 8 | 8 | 8 | 8 |
| diff_in_freq_of_2_peaks | Pearson Correlation | .111 | .044 | .140 | -.247 | .039 | -.129 | -.225 | -.553 | .093 | .604 | -.127 | -.372 | 1 | .219 |
| | Sig. (2-tailed) | .794 | .918 | .741 | .555 | .927 | .760 | .591 | .155 | .827 | .113 | .764 | .365 | | .602 |
| | N | 8 | 8 | 8 | 8 | 8 | 8 | 8 | 8 | 8 | 8 | 8 | 8 | 8 | 8 |
| diff_in_amp_of_2_peaks | Pearson Correlation | **.138** | **.167** | **.044** | **-.203** | **.057** | **-.344** | **-.286** | **-.353** | **.367** | **.135** | **.028** | **-.053** | **.219** | **1** |
| | Sig. (2-tailed) | **.744** | **.692** | **.917** | **.629** | **.894** | **.403** | **.493** | **.390** | **.371** | **.749** | **.947** | **.900** | **.602** | |
| | N | **8** | **8** | **8** | **8** | **8** | **8** | **8** | **8** | **8** | **8** | **8** | **8** | **8** | **8** |
| **. Correlation is significant at the 0.01 level (2-tailed). | | | | | | | | | | | | | | | |
| *. Correlation is significant at the 0.05 level (2-tailed). | | | | | | | | | | | | | | | |

### Factor analysis for timbre parameters

We are working with a huge number of timbre parameters but considering all the parameters simultaneously is not a good option as the parameters might be correlated among themselves. Hence to understand the underlying pattern, factor analysis has been undertaken. The essential purpose of factor analysis is to describe, if possible, the covariance relationships among many variables in terms of a few underlying, but unobservable, random quantities called factors. Suppose variables can be grouped by their correlations i.e., suppose all variables within a particular group are highly correlated among themselves but have relatively small correlations with variables in different group. Then it is conceivable that each group of variables represents a single underlying construct, or factor, that is responsible for the observed correlations and chooses a variable from each group if possible for data reduction. Here the whole data set including 14 parameters of timbre can be classified into 5 underlying factors, which can explain 93.874 % of the total variation in the dataset.

For factor analysis, we have used the varimax orthogonal rotation procedure through Principle Component Analysis and have considered five underlying factors. Factors produced in the initial extraction phase are often difficult to interpret. This is because the procedure in this phase ignores the possibility that variables identified to load on or represent factors may already have high loadings (correlations) with previous factors extracted. This may result in significant cross-loadings in which many factors are correlated with many variables. This makes interpretation of each factor difficult, because different factors are represented by the same variables. The rotation phase serves to "sharpen" the factors by identifying those variables that load on one factor and not on another. The ultimate effect of the rotation phase is to achieve a simpler, theoretically more meaningful factor pattern. The size of the factor loadings (correlation coefficients between the variables and the factors they represent) will help in the interpretation. As a general rule, variables with large loadings indicate that they are representative of the factor, while small loadings suggest that they are not. It should be kept in mind that negative factor loading implies negative correlation with the underlying factor and the other loadings. For example, in the rotated component matrix T2 has a negative loading while T3 has a positive loading (.747), which means that T2 and T3 are negatively correlated and also T2 has a negative correlation with the underlying Factor 1.

**Table 3: Descriptive Statistics**

| | Mean | Std. Deviation |
|---|---|---|
| Brightness | **12.350850** | **1.9059274** |
| Tristimulus1 | **.017563** | **.0146337** |
| Tristimulus2 | **.079800** | **.0272315** |
| Tristimulus3 | **.902487** | **.0179607** |
| Odd parameter | **.479525** | **.0171888** |
| Even Parameter | **.502688** | **.0097811** |
| Spectral Irregularity | **.134263** | **.0324775** |
| Spectral inharmonicity | **-.986013** | **2.4384462** |
| Spectral Centroid | **24.425163** | **.1749130** |
| pitch | **259.041650** | **54.3488050** |
| Attack time | **.012350** | **.0022078** |
| Average RMS power | **-12.865000** | **6.8755000** |
| Diff in freq of 2 peaks | **303.041675** | **111.4243042** |
| Diff in amp of 2 peaks | **8.029175** | **4.8769795** |

This is the table for descriptive statistics for all the timbre parameters.

**Table 4: table of communalities of timbre parameters**

| **Communalities** | | |
|---|---|---|
| | Initial | Extraction |
| Brightness | **1.000** | **.987** |
| Tristimulus1 | **1.000** | **.921** |
| Tristimulus2 | **1.000** | **.996** |
| Tristimulus3 | **1.000** | **.919** |
| Odd parameter | **1.000** | **.988** |
| Even Parameter | **1.000** | **.941** |
| Spectral Irregularity | **1.000** | **.959** |
| Spectral inharmonicity | **1.000** | **.893** |
| Spectral Centroid | **1.000** | **.891** |
| pitch | **1.000** | **.986** |
| Attack time | **1.000** | **.784** |
| Average RMS power | **1.000** | **.982** |
| Diff in freq of 2 peaks | **1.000** | **.919** |
| Diff in amp of 2 peaks | **1.000** | **.976** |
| Extraction Method: Principal Component Analysis. | | |

This is the table of communalities which shows how much of the variance in the variables has been accounted for by the extracted factors. For instance 99.6% of the variance in Tristimulus2 is accounted for while 78.4% of the variance in Attack time is accounted for.

**Table 5 : Explanation of variance**

| Total Variance Explained | | | | | | | | | |
|---|---|---|---|---|---|---|---|---|---|
| Component | Initial Eigenvalues | | | Extraction Sums of Squared Loadings | | | Rotation Sums of Squared Loadings | | |
| | Total | % of Variance | Cumulative % | Total | % of Variance | Cumulative % | Total | % of Variance | Cumulative % |
| 1 | **4.736** | **33.828** | **33.828** | **4.736** | **33.828** | **33.828** | **3.326** | **23.756** | **23.756** |
| 2 | **3.349** | **23.918** | **57.746** | **3.349** | **23.918** | **57.746** | **3.029** | **21.638** | **45.395** |
| 3 | **2.520** | **18.001** | **75.747** | **2.520** | **18.001** | **75.747** | **2.962** | **21.158** | **66.553** |
| 4 | **1.520** | **10.858** | **86.605** | **1.520** | **10.858** | **86.605** | **2.548** | **18.203** | **84.755** |
| 5 | **1.018** | **7.269** | **93.874** | **1.018** | **7.269** | **93.874** | **1.277** | **9.119** | **93.874** |
| 6 | **.546** | **3.902** | **97.776** | | | | | | |
| 7 | **.311** | **2.224** | **100.000** | | | | | | |
| 8 | **2.801E-016** | **2.001E-015** | **100.000** | | | | | | |
| 9 | **1.792E-016** | **1.280E-015** | **100.000** | | | | | | |
| 10 | **1.948E-017** | **1.392E-016** | **100.000** | | | | | | |
| 11 | **-5.120E-017** | **-3.657E-016** | **100.000** | | | | | | |
| 12 | **-7.063E-017** | **-5.045E-016** | **100.000** | | | | | | |
| 13 | **-2.223E-016** | **-1.588E-015** | **100.000** | | | | | | |
| 14 | **-4.431E-016** | **-3.165E-015** | **100.000** | | | | | | |
| Extraction Method: Principal Component Analysis. | | | | | | | | | |

The above table shows all the factors extractable from the analysis along with their eigenvalues, the percentage of variance attributable to each factor, and the cumulative variance of the factor and the previous factors. Notice that the first factor accounts for **33.828** % of the variance, the second **23.918** % of the variance, the third **18.001**%, the fourth **10.858**% and the fifth **7.269**%. All the remaining factors are not significant.

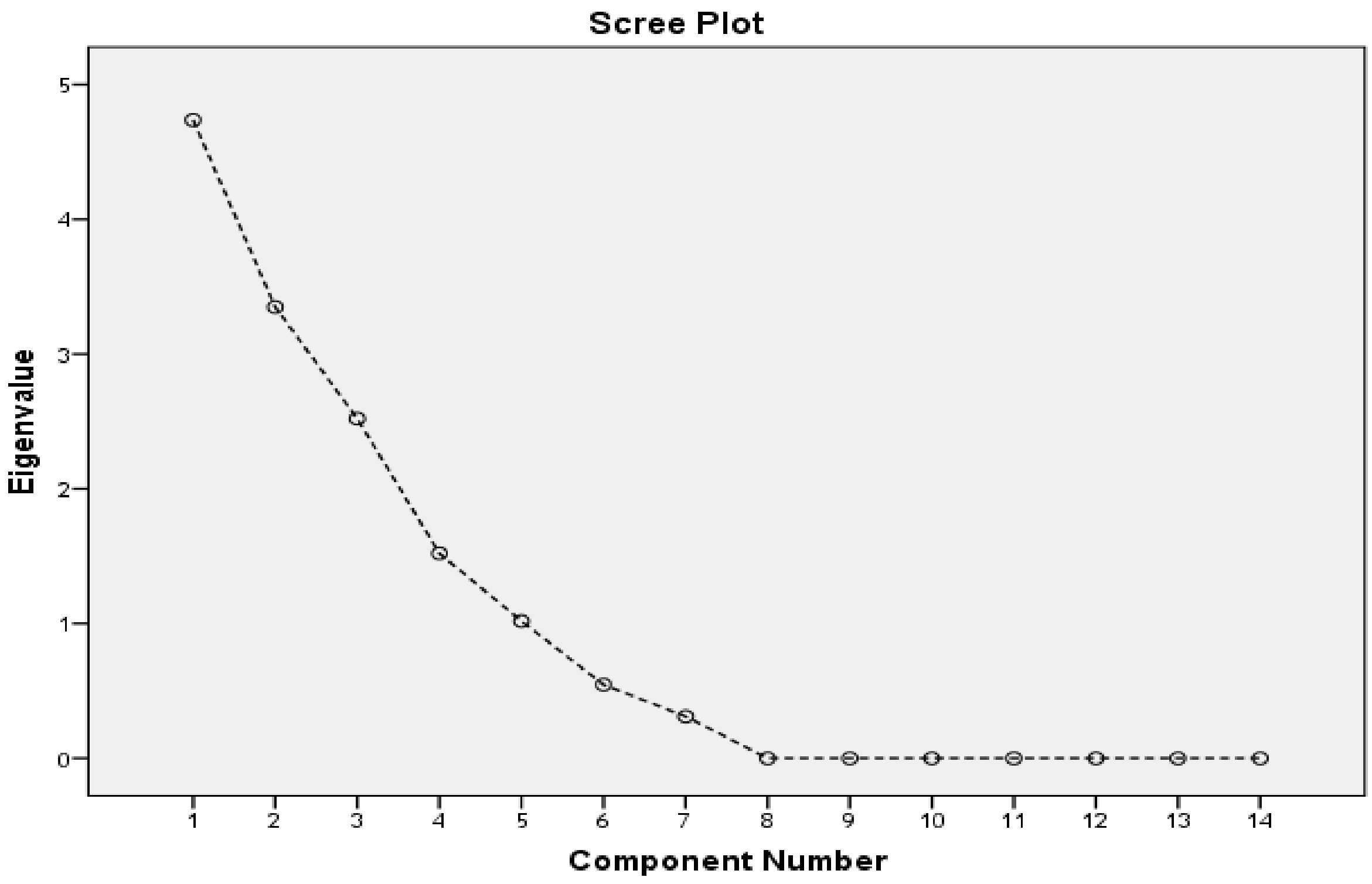

**Fig 9: Scree plot of eigenvalues**

The scree plot is a graph of the eigenvalues against all the factors. The graph is useful for determining how many factors to retain. For the first five factors the Eigen values are greater than 1 and explains significant amount of variance. Hence we take the first five factors in account for further analysis.

**Table 6: Component factor in rotated component matrix**

| Rotated Component Matrix[a] | | | | | |
|---|---|---|---|---|---|
| | Component/ Factor | | | | |
| | 1 | 2 | 3 | 4 | 5 |
| Brightness | | **-.813** | | | |
| Tristimulus1 | | | **.933** | | |
| Tristimulus2 | **-.538** | | **-.829** | | |
| Tristimulus3 | **.747** | | | | |
| Odd parameter | | | **-.857** | | |
| Even Parameter | **-.815** | | | | |
| Spectral Irregularity | | **.942** | | | |
| Spectral inharmonicity | | | | **-.842** | |
| Spectral Centroid | **.574** | **-.534** | | | |
| pitch | | | | **.899** | |
| Attack time | **.874** | | | | |
| Average RMS power | | **.959** | | | |
| diff in freq of 2 peaks | | | | **.849** | |
| Diff in amp of2 peaks | | | | | **.965** |
| Extraction Method: Principal Component Analysis. Rotation Method: Varimax with Kaiser Normalization. | | | | | |
| a. Rotation converged in 7 iterations. | | | | | |

Given below are the explanations of the rotated component matrix.

1) Factor 1 is composed of Tristimulus2, Tristimulus3, Even Parameter, Spectral Centroid, Attack time and hence these are highly inter-correlated among themselves.
2) Brightness, Spectral Irregularity, Spectral Centroid and Average RMS power have very high loading in Factor 2.
3) Tristimulus1, Tristimulus2 and Odd parameter are substantially loaded on Factor 3.
4) Factor 4 is composed of Spectral inharmonicity, pitch and difference in frequency of 2 highest peaks.
5) Difference in amplitude of 2 highest peaks is the only contributor in the Factor 5.

One can choose to name these factors based on which parameters they are made of. So we obtain the clustering pattern of the parameters and how they are correlated.

**Factor analysis of the notations**

We have carried out factor analysis on the notations. Thus we have obtained the following Correlation matrix and Component Loading matrix.

**Table 7: Correlation among notation**

| **Correlations for Notations** | | | | | | | | | |
|---|---|---|---|---|---|---|---|---|---|
| | | TA | TI | TEEN | THUN | TU | TE | RE | GHE |
| TA | Pearson Correlation | **1** | **.987**** | **.987**** | **.980**** | **.981**** | **.997**** | **.996**** | **.980**** |
| | Sig. (2-tailed) | | **.000** | **.000** | **.000** | **.000** | **.000** | **.000** | **.000** |
| | N | **14** | **14** | **14** | **14** | **14** | **14** | **14** | **14** |
| TI | Pearson Correlation | **.987**** | **1** | **.951**** | **.937**** | **.938**** | **.972**** | **.994**** | **.939**** |
| | Sig. (2-tailed) | **.000** | | **.000** | **.000** | **.000** | **.000** | **.000** | **.000** |
| | N | **14** | **14** | **14** | **14** | **14** | **14** | **14** | **14** |
| TEEN | Pearson Correlation | **.987**** | **.951**** | **1** | **.999**** | **.998**** | **.996**** | **.974**** | **.997**** |
| | Sig. (2-tailed) | **.000** | **.000** | | **.000** | **.000** | **.000** | **.000** | **.000** |
| | N | **14** | **14** | **14** | **14** | **14** | **14** | **14** | **14** |
| THUN | Pearson Correlation | **.980**** | **.937**** | **.999**** | **1** | **1.000**** | **.992**** | **.964**** | **.997**** |
| | Sig. (2-tailed) | **.000** | **.000** | **.000** | | **.000** | **.000** | **.000** | **.000** |
| | N | **14** | **14** | **14** | **14** | **14** | **14** | **14** | **14** |
| TU | Pearson Correlation | **.981**** | **.938**** | **.998**** | **1.000**** | **1** | **.992**** | **.964**** | **.997**** |
| | Sig. (2-tailed) | **.000** | **.000** | **.000** | **.000** | | **.000** | **.000** | **.000** |
| | N | **14** | **14** | **14** | **14** | **14** | **14** | **14** | **14** |
| TE | Pearson Correlation | **.997**** | **.972**** | **.996**** | **.992**** | **.992**** | **1** | **.989**** | **.992**** |
| | Sig. (2-tailed) | **.000** | **.000** | **.000** | **.000** | **.000** | | **.000** | **.000** |
| | N | **14** | **14** | **14** | **14** | **14** | **14** | **14** | **14** |
| RE | Pearson Correlation | .996** | .994** | .974** | .964** | .964** | .989** | 1 | .969** |
| | Sig. (2-tailed) | .000 | .000 | .000 | .000 | .000 | .000 | | .000 |
| | N | 14 | 14 | 14 | 14 | 14 | 14 | 14 | 14 |
| GHE | Pearson Correlation | .980** | .939** | .997** | .997** | .997** | .992** | .969** | 1 |
| | Sig. (2-tailed) | .000 | .000 | .000 | .000 | .000 | .000 | .000 | |
| | N | 14 | 14 | 14 | 14 | 14 | 14 | 14 | 14 |

**. Correlation is significant at the 0.01 level (2-tailed).

We find that the all the notations are highly positively correlated among themselves.

**Table 8: Component analysis for strokes**

| Component Matrix[a] | |
|---|---|
| | Component |
| | 1 |
| TA | **.997** |
| TEE | **.973** |
| TEEN | **.996** |
| DIN | **.992** |
| THU | **.992** |
| TE | **1.000** |
| RE | **.990** |
| GHE | **.992** |

Extraction Method: Principal Component Analysis.

a. 1 components extracted.

Since the notations are highly correlated only 1 Factor/Component can be extracted which explains 98.316% of the total variability of the whole dataset.

**Conclusion**

(i) The two drums of a tabla produce many different timbres. Tabla strokes have unique harmonic and timbral characteristics at mid frequency range and have no uniqueness at low frequency ranges. These provide to facilitate the development and transmission of a sophisticated solo repertoire. In addition to the rhythmic complexity of tabla music, it is its timbral beauty and diversity that distinguish it from other percussion instruments.

(ii) In general strokes made at the vicinity of centre circle pumped up energy at high frequency range and they are the brightest stokes while the strokes made at the edge of the membrane are weakest strokes having low energy.

(iii) Change of timbre parameters of strokes of same tabla occurs due to difference in stroke execution.

(iv) Dimention of tabla viz. diameter of the membrane, dimension of hollow chamber etc are the determining factor of tabla timbre.

(v) Statistical analysis shows that timbre parameters have 5 underlying factors which explain the total variability in the data set but notations are highly correlated among themselves and hence there is only 1 underlying factor.

(vi) Tristimulus2, Tristimulus3, Even Parameter, Spectral Centroid, Attack time are the most important timbre features to study timbre of tabla followed by Brightness, Spectral Irregularity, Spectral Centroid, Average RMS power based on the total variability explained.

(vii) Tristimulus2, Tristimulus3, Even Parameter, Spectral Centroid and Attack time are the most important timbre parameters to explain tabla strokes, since they have highest degree of variance and eigen values.

(viii) Brightness, Spectral Irregularity, Spectral Centroid and Average RMS power have very high loading and are also important timbre parameters.

(ix) Tristimulus1, Tristimulus2 and Odd parameter have low variance and eigen values, so may be used little to explain tabla strokes.

(x) Other timbre parameters are not significant in explaining tabla strokes.

**References**

1. Anirban Patranabis, Kaushik Banerjee, Ranjan Sengupta and Dipak Ghosh, '*Objective Analysis of the Timbral quality of Sitars having Structural change over Time*', Ninaad (Journal of ITC Sangeet Research Academy), Vol. 25, pp 1-7, 2011

2. B M Banerjee and D Nag, "*The Acoustical Character of Sounds from Indian Twin Drums*", ACUSTICA, Vol.75, 206-208, (1991)

3. B S Ramakrishna "*Modes of Vibration of the Indian Drum Dugga or left hand Thabala*", J. Acoust. Soc. Am. Vol.29, 234-238, (1957)

4. C V Raman "*The Indian Musical Drums*", Proc. Indian Acad. Sc., 1A, pp. 179-188, (1934)

5. C V Raman and Sivakali Kumar "*Musical drums with harmonic overtones*", Nature, Vol.104, 500, (1920)

6. C. Valens, "*A Really Friendly Guide to Wavelets (C)*", (1999-2004), wavelets@polyvalens.com

7. David Courtney "*Psychoacoustics of the Musical pitch of Tabla*" Journal of SanGeet Research Academy, Calcutta, India, Vol.13, No 1, (1999)

8. Grey J. M. "*Multidimensional Perceptual Scaling of Musical Timbre*", Journal of the Acoustical Society of America Vol. 61, 1270 – 1277. (1977)

9. K N Rao "*Theory of the Indian Musical Drums*", Proc. Indian Acad. Sci. Vol.7A, 75-84, (1938)

10. R N Ghosh "*Note on Musical Drums*", Phys. Review, Vol.20, 526-527, (1922)

11. Robi Polikar, '*The Wavelet Tutorial*', Rowan University http://users.rowan.edu/polikar/wavelets/wttutorial.html

12. Sengupta R, Dey N, Nag D, Datta A K and Parui S K (2004), *Objective Evaluation of Tanpura from the Sound Signals using Spectral Features*‖. Journal of ITC Sangeet Research Academy, vol.18

13. Tae Hong Park, "*Towards Automatic Musical Instrument Timbre Recognition*", PhD thesis, Department of Music, Princeton University, (2004)

14. T Sarojini and A Rahman "*Variational Method for the Vibrations of Indian Drums*" J. Acoust. Soc. Am. Vol.30, 191-196, (1958)